\documentclass[aps,prl,twocolumn,superscriptaddress,showpacs,floatfix]{revtex4}

\usepackage{graphicx}
\usepackage{longtable}

\begin{document}

\title{Collective Shape-Phases of Interacting Fermion Systems}

\author{Yu-xin Liu}
\affiliation{Department of Physics and the CMOE Key Laboratory of
Heavy Ion Physics, Peking University, Beijing 100871, China}
\affiliation{Center of Theoretical Nuclear Physics, National
Laboratory of Heavy Ion Accelerator, Lanzhou 730000, China}

\author{Zhan-feng Hou}
\affiliation{Department of Physics and the CMOE Key Laboratory of
Heavy Ion Physics, Peking University, Beijing 100871, China}

\author{Yu Zhang}
\affiliation{Department of Physics and the CMOE Key Laboratory of
Heavy Ion Physics, Peking University, Beijing 100871, China}

\author{Haiqing Wei}
\affiliation{School of Information Science and Engineering, Lanzhou
University, Lanzhou 730000, China}

\date{\today}

\begin{abstract}
A microscopic theory is presented for identifying shape-phase
structures and transitions in interacting fermion systems. The
method provides a microscopic description for collective
shape-phases, and reveals detailed dependence of such shape-phases
on microscopic interaction strengths. The theory is generally
applicable to fermion systems such as nuclei, quarks, and in
particular trapped cold atoms, where shape-phases may be observed
and investigated in a controlled manner.
\end{abstract}

\pacs{21.10.-k, 21.60.-n, 21.60.Cs, 21.60.Fw}

\maketitle

\newpage

\parindent=20pt

%\section{Introduction}

Collective motion is common in fermion systems such as nuclei and
molecules. Nuclei have been found for a long time to possess
interesting geometric shapes, such as a vibrating spheroid, an
ellipsoid, and exotically deformed shapes~\cite{BM75,Heenen05}.
Pairing induced deformations (PIDs) have recently been observed in
trapped fermion systems of cold atoms~\cite{Zwierlein06,Hulet06}.
Mean-field theories have been used to describe the properties of
such fermion systems as many-electron atoms, nuclei, quarks, and
trapped atomic fermions~\cite{Lawson80Heyde90,Liu03prl,CR05}. A
long-standing question is to identify the shape phase structures
and transitions in the general framework of mean-field theory in
fermion space. Only recently there have been studies on nuclear
shape-phase transitions~\cite{Rowe98,Luo06} and critical point
symmetries~\cite{Ginocchio05} in the framework of shell model
having either monopole-pairing and quadrupole-quadrupole (QQ)
interactions or monopole-pairing and quadrupole-pairing and on the
shape coexistence and evolution in superheavy nuclei in the energy
density functional theory involving monopole-pairing
term~\cite{Heenen05}. But the overall structure of nuclear
shape-phases is still unclear, especially in a more general
mean-field model that incorporates not only monopole-pairing and
QQ interactions but also one-body terms and quadrupole-pairing. In
similar systems of trapped atoms, there is also a lack of
theoretical understanding of PIDs observed in recent
experiments~\cite{Zwierlein06,Hulet06}.

Starting from a microscopic theory of interacting fermion systems,
this paper presents a systematic method for identifying
shape-phase structures and transitions in such systems, so to
provide a unified microscopic description for collective
shape-phases of both nuclei and trapped cold atoms, as well as
other similar systems. From a microscopic viewpoint, our method
recovers the rich structures and transitions of the shape-phases
(dynamical symmetries) that have been experimentally observed and
accounted for by the interacting boson model (IBM)
\cite{CoherentState,Iachello87}. More importantly, our method
reveals detailed dependence of such shape-phases on the
interaction strengths in a mean-field Hamiltonian. The same theory
may be applied to other systems such as trapped cold atoms with
tunable interaction parameters, therefore points to the
possibility of observing and controlling shape-phases and
transitions therein.

As an approximation of the shell model, a successful mean-field
theory of nuclear structure, the IBM has been proven highly
successful in describing low-lying collective features of medium
and heavy mass nuclei \cite{IachelloRMP,Iachello87}. In
particular, it has well accounted for nuclear shape-phase
structures and transitions \cite{CoherentState,Iachello87,Jolie01,
Leviatan03,Iachello04,Cejn034, Pan03,Rowe045,Liu06}). The success
can be attributed to the incorporation of collective degrees of
freedom and the simplification of calculations by a mapping from
fermions to bosons. Three dynamical symmetries, U(5), SU(3) and
O(6), are naturally incorporated into the IBM, which correspond
respectively to shape-phases of a spheroid, axially prolate rotor
and $\gamma$-soft rotors \cite{CoherentState,Iachello87}. There is
also $\overline{\mbox{SU(3)}}$ symmetry corresponding to an
axially oblate rotor phase\cite{Iachello87}. A widely used
mean-field Hamiltonian in fermion space can be written
as\cite{Higashiyama02}
\begin{equation}
  \hat{H}_F = % \hat{H}_0 + \hat{H}_{0P} + \hat{H}_{2P} + \hat{H}_{QQ}\ ,
             \sum_{jm} \varepsilon_{j} a_{jm}^{\dag} a_{jm}
             -\frac{1}{2} g_0 \hat{P}_0^{\dag} \hat{P}_0
             -\frac{1}{2} g_2 \hat{P}_{2}^{\dag} \cdot \hat{P}_{2}
             -\frac{1}{2} k :\! \hat{Q}_{2}\cdot \hat{Q}_{2}\!: \, ,
\end{equation}
with
\begin{eqnarray}
  \hat{P}_0^{\dag} & = & \sum_{j m} a_{j m}^{\dag} \tilde{a}_{j m}^{\dag}\ ,\\
  \hat{P}_{2\mu}^{\dag} & = & \sum_{j_1 m_1 j_2 m_2}<j_1m_1|q_{2\mu}|j_2m_2> a_{j_1
m_1}^{\dag} \tilde{a}_{j_2 m_2}^{\dag}\ ,\\
  \hat{Q}_{2\mu} & = & \sum_{j_1 m_1 j_2 m_2}<j_1m_1|q_{2\mu}|j_2m_2> a_{j_1
m_1}^{\dag} a_{j_2 m_2}\ ,
\end{eqnarray}
where $\varepsilon_{j}$ is the single-particle energy, $:\ :$
denotes the normal product of fermion operators, $g_{0}$, $g_{2}$,
$k$ are strengths of monopole-pairing, quadrupole-pairing, and QQ
interaction, respectively, $\tilde{a}_{j m}=(-1)^{j-m} a_{j -m}$
and $q_{2\mu}=r^2 Y_{2\mu}$. Besides the energy spectrum, a
particularly interested property is the electric quadrupole
transition rate, $B(E2,L_{i}\rightarrow L_{f})=
\frac{1}{2L_{i}+1}\langle L_{f} || \hat{T}(E2) || L_{i} \rangle
^{2} $, with $\hat{T}(E2)$ being $\hat{Q}_{2}$ multiplied by an
effective charge.

We take a nuclear system as a numerical example, although the same
theory and numerical procedure can be applied to other fermion
systems equally well. To simplify the calculations, the method of
Dyson-type boson mapping \cite{Dyson,Klein} is employed with the
$SD$ pair truncation keeping only pairs with angular momentum
$J=0$ and $J=2$ \cite{SDPair}. Then the Hamiltonian takes an IBM
form $\hat{H}_B(S^{\dag}, S, D^{\dag}, D )$, with the $S$, $D$
pairs being regarded as $s$, $d$ bosons, respectively. Protons and
neutrons are not differentiated, so the model is actually
corresponding to the IBM-1. The single-particle wave functions are
chosen to be harmonic oscillator's with oscillation constant
$b^2=1.0A^{1/3} fm^{2}$, $A=130$. For the 50-82 shells, the active
single-particle orbits are $2d_{5/2}, 1g_{7/2}, 3s_{1/2},
1h_{11/2}, 2d_{3/2}$ with energies $0, 0.8, 1.3, 2.5, 2.8$ MeV,
respectively, similar to those used in Ref.~\cite{Uher}.  The
final truncated Hamiltonian is diagonalized in the U(5)-symmetric
bases with total number of nucleon pairs being set to $N=10$.

To identify the shape-phases, it is helpful to examine the
correspondence between the interaction strengthes in the
microscopic model and the dynamical symmetries in the IBM.
Quantities of interest are the energy ratios $R_{42} =
\frac{E_{4_1}}{E_{2_1}}$, $R_{62} = \frac{E_{6_1}}{E_{2_1}}$,
$R_{02} = \frac{E_{0_2}}{E_{2_1}}$, and $R_{22} =
\frac{E_{2_2}}{E_{2_1}}$, with $E_{0_1}=0$, and ratios of the
electric quadrupole transition rates $B_{42}=\frac{B(E2;4_1
\rightarrow 2_1)}{B(E2;2_1 \rightarrow 0_1)}$,
$B_{64}=\frac{B(E2;6_1 \rightarrow 4_1)}{B(E2;2_1 \rightarrow
0_1)}$, $B_{02}=\frac{B(E2;0_2 \rightarrow 2_1)} {B(E2;2_1
\rightarrow 0_1)}$, and $B_{22}= \frac{B(E2;2_2 \rightarrow
2_1)}{B(E2;2_1 \rightarrow 0_1)}$, which are known to be able to
characterize the low-lying energy spectrum well. Table~\ref{t1}
lists values of these quantities in the dynamical symmetries of
IBM-1 (with total boson number $N=10$), which are to be compared
with those obtained in the microscopic model.

\begin{table}[ht]
\caption{\label{t1} Values of interested quantities of a 10-boson
system in the IBM. Those marked with a star depend on additional
parameters in the Hamiltonian (as given in Refs.~\cite{Rowe045,
Iachello04}).} \vspace*{-6pt}
%with $ \hat{H} = (1-\alpha) \hat{n}_d + \frac{\alpha}{4N}
%\hat{Q}(\chi)\cdot \hat{Q}(\chi)$
\begin{center}
\begin{tabular}{l|cccccccc}
\hline\hline
      & \ $R_{42}$\ & \ $R_{62}$ \  & $R_{02}$\ & \ $R_{22}$ \  & \ $B_{42}$ \ & \ $B_{64}$ \
      & \ $B_{02}$\ & \ $B_{22}$ \  \\
\hline
 U(5) &\ 2.00\ &\ 3.00\ &\ 2.00$^*$  &\ 2.00$^*$  & \ 1.80\ &\ 2.40\ &\ 1.80\ &\ 1.80 \\
 O(6) &\ 2.50\ &\ 4.50\ &\ 4.50$^*$  &\ 2.50$^*$  & \ 1.38\ &\ 1.52\ &\ 0.00\ &\ 1.38 \\
 SU(3)&\ 3.33\ &\ 7.00\ &\ 23.7$^*$  &\ 24.7$^*$  & \ 1.40\ &\ 1.48\ &\ 0.00\ &\ 0.00 \\
\hline\hline
\end{tabular}
\end{center}
\end{table}

\begin{figure}
%\begin{center}     %\centering \scalebox{0.610}{
\includegraphics[scale=0.5,angle=0]{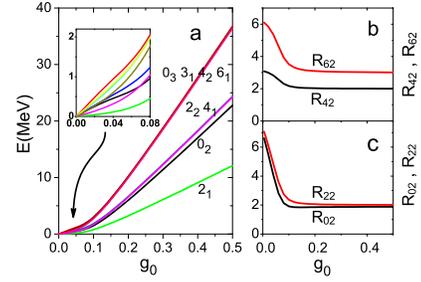}     %[bb=15 10 350 240]{eg0.eps}}
\caption{The dependence of low-lying levels on $g_0$ when
$g_2=k=0$.} \label{feg0}
%\end{center}
\end{figure}

\begin{figure}
%\begin{center}   %ing \scalebox{0.610}{
\includegraphics[scale=0.5,angle=0]{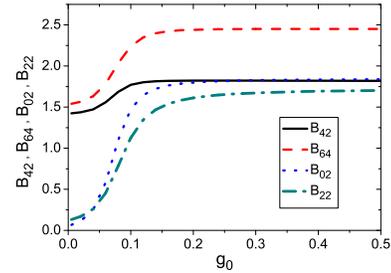}
\caption{ The dependence of $B_{42}$, $B_{64}$, $B_{02}$, and
$B_{22}$ on $g_{0}$ when $g_2=k=0$.} \label{fbg0}
%\end{center}
\end{figure}

We first look at the effect of monopole-pairing with parameters
$g_{0} \in [0, 0.50]$ and $g_{2} = k =0$. Figs.~\ref{feg0} and
\ref{fbg0} show the dependence of the energy levels and the B(E2)
ratios on $g_0$. The degenerate U(5) levels, such as the $2_2-4_1$
doublet and the $0_3-3_1-4_2-6_1$ quartet, are well reproduced at
$g_{0} > 0.12$, and the energy and $B(E2)$ ratios are
U(5)-symmetrically valued. So a large $g_0$ favors a spherical
phase. The critical value of $g_0\approx 0.12$ agrees well with
the empirical value $g_0 \sim 20/A$ with $A=130$ assigned in
Ref.~\cite{Bes69}. In addition, as $g_{0}$ is very small, the
calculation gives approximately the feature of the SU(3) or
$\overline{\mbox{SU(3)}}$ symmetry. It indicates that a fermion
system with very weak monopole-pairing may appear in an axially
deformed shape.

\begin{figure}
%\begin{center}      %\centering \scalebox{0.610}{
\includegraphics[scale=0.5,angle=0]{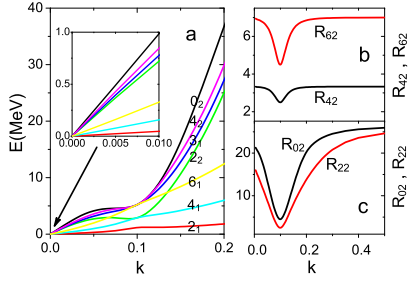}    %[bb=10 15 350 225]{ek.eps}}
\caption{The calculated energy levels as functions of $k$ when
$g_0=g_2=0$.} \label{fek}
%\end{center}
\end{figure}

\begin{figure}
%\begin{center}                % \centering \scalebox{0.610}{
\includegraphics[scale=0.5,angle=0]{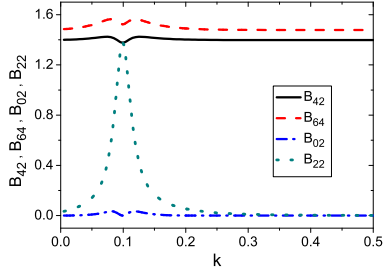}    % [bb=10 15 350 225]{bk.eps}}
\caption{ The calculated $B(E2)$ ratios $B_{42}$, $B_{64}$,
$B_{02}$, and $B_{22}$ as functions of $k$ when $g_0=g_2=0$.}
\label{fbk}
%\end{center}
\end{figure}

Next we examine the effects of the QQ interaction, which is known to
coincide with Elliott's model \cite{Elliott58}, where the $Q_2$ is a
quadrupole tensor in the SU(3) symmetry. However, a system may not
be limited to the SU(3) phase in a more realistic model with
one-body terms. So we did calculations with parameters
$k\in[0,0.50]$ and $g_0=g_2=0$. The energy levels and the $B(E2)$
ratios as functions of $k$ are shown in Figs.~\ref{fek} and
\ref{fbk} respectively, where $k$ is limited to below $0.20$ in part
(a) of Fig.~\ref{fek} for a better view of the energy levels. There
is clearly a special point $k\approx 0.10$, at which all the
quantities defined in Table~\ref{t1} are approximately symmetric.
More specifically, this point corresponds to an O(6) phase with
$R_{42}\approx R_{22}\approx 2.5$, $R_{62}\approx R_{02}\approx
4.5$, $B_{42}\approx 1.38$, $B_{64}\approx 1.52$, $B_{02} \approx
0$, and $B_{42}\approx 1.38$, in excellent agreement with the
corresponding values in Table~\ref{t1}. In addition, the degeneracy
of the $2_2-4_1$ doublet and the $0_2-3_1-4_2-6_1$ quartet is
clearly seen in part (a) of Fig.~\ref{fek}. On the other hand, at
the two extremes of $k<0.005$ and $k\geq 0.2$, the axially symmetric
rotation phase prevails and the calculated energies and $B(E2)$
ratios well reproduce those of the SU(3) symmetry in Table~\ref{t1}.
%(for example, $R_{42} = 3.3$, $R_{62} = 6.9$, $R_{02} = 21$, $R_{22}
%= 16$, $B_{42} = 1.4 $, $B_{64} = 1.5 $, $B_{02} = 0.00$ and $B_{22}
%<0.05$).

It is known that an axially symmetric rotor could have either a
prolate or an oblate shape, corresponding to the SU(3) or
$\overline{\mbox{SU(3)}}$ phase in IBM-1. Since the parameters in
Table~\ref{t1} are the same for both phases, we have to implement
other quantities to distinguish them. One parameter serving this
purpose is the quadrupole moment $Q(2_1^+)$
%$=\frac{4}{5}\sqrt{\pi}\langle2\ 2\ 2\ 0 \vert 2\ 2\rangle \langle
%2_1^+ \Vert \hat{T}(E2) \Vert 2_1^+ \rangle$,
%
which is negative-, zero-, and positive-valued in the SU(3), O(6),
and $\overline{\mbox{SU(3)}}$ phases respectively \cite{Jolie01}.
As the QQ interaction strength $k$ increases from below to above
$0.1$, the calculated $Q(2_1^+)$ (with the effective charge
normalized to $1$) decreases from a positive value to zero and
then to negative, which indicates an evolution of symmetry
$\overline{\mbox{SU(3)}}\rightarrow$ O(6) $\rightarrow$ SU(3),
namely, transitions from oblate to prolate shape with a critical
point of $\gamma$-soft rotation in between. This agrees
excellently with the variation behavior against the parameter
$\chi$ in Ref.~\cite{Jolie01}. In addition, reexamining the effect
of the $g_{0}$, we know that, as $g_{0}$ is very small, the
$Q(2_1^+)$ is positive. It manifests that the fermion system with
weak monopole-pairing takes an oblate shape, {\it i.e.}, it
appears as a plateau. Such a result reproduces the recently
observed density distribution of trapped cold
atoms\cite{Zwierlein06,Hulet06} qualitatively.

%\begin{figure}
%\centering \scalebox{0.7}{
%\includegraphics[bb=10 15 275 225]{qk.eps}}
%\caption{ The calculated values of $Q(2_1^+)$. The values of $g_0$,
%$g_2$ and $k$ are the same as that in Fig.~\ref{fek}. }\label{fqk}
%\end{figure}

\begin{figure}
%\begin{center}     %\centering \scalebox{0.610}{
\includegraphics[scale=0.5,angle=0]{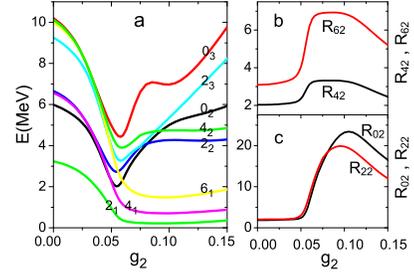}     % [bb=10 15 350 225]{eg2.eps}}
\caption{The dependence of low-lying levels on $g_2$ when $g_0=0.15$
and $k=0$.} \label{feg2}
%\end{center}
\end{figure}

\begin{figure}
%\begin{center}    % \centering \scalebox{0.610}{
\includegraphics[scale=0.5,angle=0]{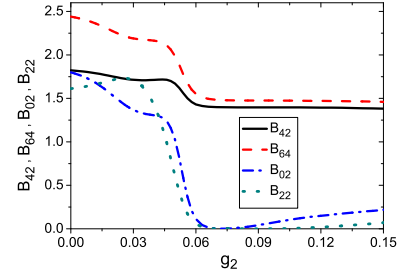}    % [bb=10 15 350 235]{bg2.eps}}
\caption{The dependence of $B_{42}$, $B_{64}$, $B_{02}$, and
$B_{22}$ on $g_{2}$ when $g_0=0.15$ and $k=0$.} \label{fbg2}
%\end{center}
\end{figure}

Finally, we study how quadrupole-pairing affects the shape-phase
structure, using parameters $g_2 \in [0, 0.15]$, $g_0=0.15$, $k=0$.
Note that $g_2<g_0$ holds true for most realistic depictions of
nuclei \cite{IachelloRMP}. The calculated dependence of low-lying
levels and $B(E2)$ ratios on the quadrupole-pairing strength is
shown in Figs.~\ref{feg2} and \ref{fbg2} respectively. For $g_2\in
[0,0.04]$, the system is evidently in a vibration phase. For $g_2\in
[0.07,0.10]$, it is in an axially symmetric rotation phase, with the
corresponding parameters in Table~\ref{t1} well reproduced.
Furthermore, $Q(2_1^+)$ is negative in this region. It shows that
the $g_2\in[0.07,0.10]$ generates an axially prolate deformation (in
SU(3) symmetry). Besides, the $g_2\in[0.04,0.07]$ induces a
U(5)-SU(3) transition, and the $g_{0} > 0.10$ suppresses collective
motion.
%
%$R_{42} \approx 3.3$, $R_{62} \approx 6.9$, $14\leq R_{02} \leq 20$,
%$15\leq R_{22} \leq 20$, $B_{42} \!\approx\! 1.4$, $B_{64}
%\!\approx\! 1.5$, $B_{02} \! < \! 0.08$ and $B_{22} \!\approx\!
%0.00$.

The above analyses may be summarized in Table~\ref{t2} for the
correspondence between regions of interaction strengths in our
microscopic model and the dynamical symmetries ({\it i.e.},
shape-phases) in the IBM. The Table indicates that the occurrence
of an axially oblate rotor should be rare in nuclei, as it is
inaccessible to most of the interactions. But it may be common in
trapped cold atomic systems since the monopole-pairing there is
very week. It may be noted that, calculations using different
configuration parameters may result in different regions and
critical values of interaction strengths corresponding to various
shape-phases, but the general correspondence between regions of
interaction strengths and shape-phases and the overall shape-phase
structure should remain for the characteristic features of an
interacting fermion system.

\begin{table}[ht]
\caption{\label{t2} The correspondence between microscopic
parameter settings and shape-phases (dynamical symmetries).}
\vspace*{-6pt}
\begin{center}
\begin{tabular}{l|cccc}
\hline\hline
& \ \ \ U(5)\ \ & \ \ SU(3)\ \  & \ O(6)\  &\ $\overline{SU(3)}$\ \\
\hline
$g_0$ (with $g_2=k=0$)         &  $> 0.12$  &  ---          & ---      & very small      \\
$g_2$(with $g_0\!=\!0.15$, $k\!=\!0$) &  $[0, 0.04]$  &  $[0.07,0.10]$ & ---   & ---     \\
$k$ (with $g_0=g_2=0$)         &  ---          & $\geq 0.20 $ & $0.10$ &  $\approx 0.0$  \\
\hline\hline
\end{tabular}
\end{center}
\end{table}

%\section{Summary and discussions}

In conclusion, we have presented a microscopic method of
interacting fermion systems and studied the dependence of their
shape-phases on the strengthes of basic interactions. Our study
yields a complete shape-phase structure of such systems by
providing an extended Casten triangle exhibiting modes of
collective vibrations, axially prolate, oblate, and $\gamma$-soft
rotations, corresponding to different regions of monopole-pairing,
quadrupole-pairing, and QQ interaction strengthes. Specifically,
strong or weak monopole-pairing leads to a vibration or an axially
oblate deformation, respectively. While quadrupole-pairing and QQ
interactions also induce rotations (i.e., deformations). In
detail, a continuous increase of the QQ strength sees oblate,
$\gamma-$soft, and prolate rotors subsequently. With a fixed
strength of vibration-corresponding monopole-pairing, a suitable
quadrupole-pairing strength may induce a transition from spherical
to axially prolate elliptical phases. But an exceedingly large
quadrupole-pairing strength suppresses collective motions. Other
than nuclei, the theory should be applicable to general fermion
systems such as trapped atomic fermions, where the interaction
strengths may be conveniently tunable. In particular, our
calculations indicate that an axially deformed shape may be
induced in trapped atomic systems by either a weak
monopole-pairing or a suitable quadrupole-pairing interaction,
then the recently observed PIDs and nonuniform density
distribution in trapped atomic systems\cite{Zwierlein06,Hulet06}
could be understood as the effect of the deformation. Besides, the
quadrupole-pairing could also be a trigger to a normal to LOFF
phase transition in fermion systems because the $D$-pair is just
the polarized (i.e., inhomogeneous) pair.

\bigskip

\begin{acknowledgments}
This work was supported by the National Natural Science Foundation
of China under contract Nos. 10425521 and 10575004,
%the Major State Basic Research Development Program under contract No. G2000077400,
the Key Grant Project of Chinese Ministry of Education (CMOE)
under contact No. 305001, and the Research Fund for the Doctoral
Program of Higher Education of China under grant No. 20040001010.
One of the authors (YXL) thanks also the support of the Foundation
for University Key Teacher by the CMOE.

\end{acknowledgments}

%\newpage

%\bibliography{reference}

\end{document}